# GUI system for Elders/Patients in Intensive Care

J. L. Raheja, Dhiraj
Machine Vision Lab
CEERI Pilani, India
{jagdish, dhiraj}@ceeri.ernet.in

D. Gopinath
Dept. of Computer Science
BITS Pilani, India
deepakg92@gmail.com

Ankit Chaudhary
Member IEEE
Dept. of Computer Science
MUM, IA USA
dr.ankit@ieee.org

*Abstract*— **In the old age, few people need special care if they are suffering from specific diseases as they can get stroke while they are in normal life routine. Also patients of any age, who are not able to walk, need to be taken care of personally but for this, either they have to be in hospital or someone like nurse should be with them for better care. This is costly in terms of money and man power. A person is needed for 24x7 care of these people. To help in this aspect we purposes a vision based system which will take input from the patient and will provide information to the specified person, who is currently may not in the patient room. This will reduce the need of man power, also a continuous monitoring would not be needed. The system is using MS Kinect for gesture detection for better accuracy and this system can be installed at home or hospital easily. The system provides GUI for simple usage and gives visual and audio feedback to user. This system work on natural hand interaction and need no training before using and also no need to wear any glove or color strip.**

I. INTRODUCTION

Elder/sick people (ESP) who are not able to talk or walk, they are dependent on other humans and need continuous monitoring. Also few patients have extreme problems when they get stroke. These people need someone to be nearby to take care of them at the time of stroke. Many times no one is around, so they face a lot of problem. In the case of heart stroke or migraine, the patient is not able to call on phone even if he wishes. In such cases, a system is needed which can call other people on a simple and short action, like a 'gesture call'. Our system provides flexibility to ESP to announce his need to their caretaker by just showing a particular gesture with the developed system. This system is targeted for the old or sick people, who are not able to express their feelings by words or they can't walk. Generally these people are in hospital or in home under continuous human monitoring.

It is possible that the person who is monitoring is not nearby as he can also has some work and can go out from the room. It is usually happen in hospitals where one hospital staff needs to take care of many patients which can be in different rooms. During this time if the ESP wants to eat something or want to call someone for help, then they are unable to do that. Our developed system helps the ESP to express their wish or need in predefined lingual description to a specified place. This place can be hospital staff room or living room in home, where generally people stay. The usage of this system is very easy as the person is sick or not able to walk and he would not be able to do complex operations.

There are many patient care software available in market these days but most of them are for the use of doctor/ hospital staff. The available patient care system like *Sunrise Critical care, Critical Care Assessment*, *picis*, *Infinium* etc, they allow hospital staff to keep track to patient condition by storing daily health parameters but nothing for the patient. Software like *Amcom* put sensors on patient body and if their condition is below threshold it sends SMS to a nurse, located in hospital. It also has a button on the device which patient can press to call to nurse and nurse has to go to the patient to check the reason. While in our system, it gives specific information to hospital staff, what the patient need. The emphasis in the system design is on easy usage and simplicity. Here, it is need not to mention that a real time response time is expected from the system as it monitoring critical patients. The flow diagram of the system working is shown in figure 1.

## II. BACKGROUND

Gesture recognition is an established area now but many applications are yet to find. Mitra [1] defines gesture recognition a process where user made gesture and receiver recognized them. Many Researchers have done excellent work in this area. Ahn [2] have developed augmented interface table using infrared cameras for pervasive environment. Initially researchers have used gloves, color strips or shirt color [3-6]. Gesture recognition is the phase in which the data analyzed from the visual images of gestures is recognized as a specific gesture. Identification of a hand gesture can be done in many ways depending on the problem to be solved [6]. There has been a lot of work with natural hand also with single camera and with multiple camera [1][7-8]. Natural hand makes the communication as easy as user is talking with other human.

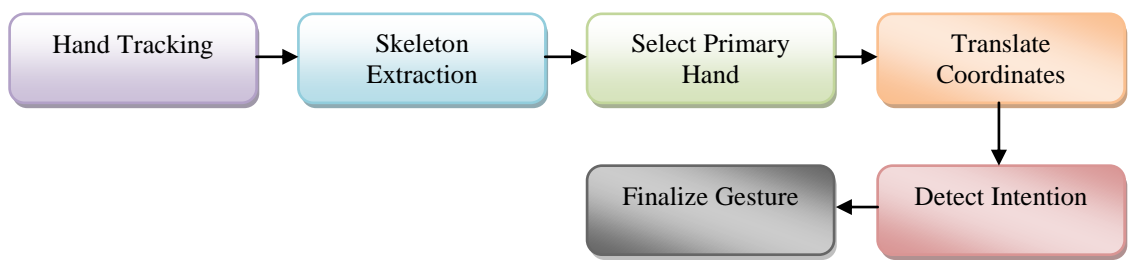

Figure. 1. Flow diagram of system working

In our previous work, we have developed a gesture based hospital monitoring system using webcam [9]. The system is available with six different options and controlled by different gesture. There was no GUI and user has to remember all the options, although it was configurable [10]. Another issue with the system was that it needs a good intensity light to recognize gesture as most of color segmentation techniques rely on histogram matching or employ a simple look-up table approach [11-12] based on the training data for the skin. The major drawback of color based localization techniques is the variability of the skin color footprint in different lighting conditions although we tried to make our system light intensity changes invariant [13] and gesture direction invariant [8]. Also in critical situations, where weather is bad or person is outside or there is no light in room like in night, the patient still can need something. So we implement with MS Kinect [14]. As with our experience, we know that it work with depth information and can provide an excellent accuracy without any light effect [15].

Although Kinect was developed mainly for games, but the technology has been applied to real-world applications as diverse as digital signage, virtual shopping, education, tele-health service delivery and other areas of health IT. Frati [16] has shown how to calculate the distance using the Kinect in 3D space. The depth images and RGB image of the object could be getting at the same time. This 3D scanner system called *Light Coding* which employs a variant of image-based 3D reconstruction. Few researchers have used Kinect to track hand parameters too [15][17].

## III. SYSTEM ARCHITECTURE

The system has a GUI which can be used either with a PC or on tablet. The User Interface of the application was developed using MS Visual Studio Designer in the Xaml file format. The canvas includes a gridbox to hold the nine option buttons, a heading label, four checkboxes to select communication option, a drop down box to indicate the proximity of the user to the Kinect sensor and radio buttons to choose left/right hand for tracking. A separate hand image is included in the system to replace the usual mouse pointer for simplicity and to make the application user friendly. The software is applicable in places like hospital wards where patients, through simple hand actions can use this.

*GUI Menu*

The GUI is developed with nine options. It is configurable and more options can be added. The 3x3 grid on the screen displays the following options, as shown in figure 2.

1. Doctor
2. Family
3. Fruits
4. Nurse
5. Emergency
6. Food
7. Bathroom
8. Water
9. Medicine

*Communication Options*

There are four ways in which the information could be sent to hospital staff. They can choose one or more than one if needed. The methods by which the user can communicate are:

1. Phone
2. Email
3. SMS
4. Voice

*Choosing the proximity*

Flexibility in terms of the proximity of the user from the Kinect sensor is also provided as a dropdown menu. When the user is farther than 3m from the Kinect, the FAR option from the dropdown menu should be selected. In all other cases, the NEAR option must be selected.

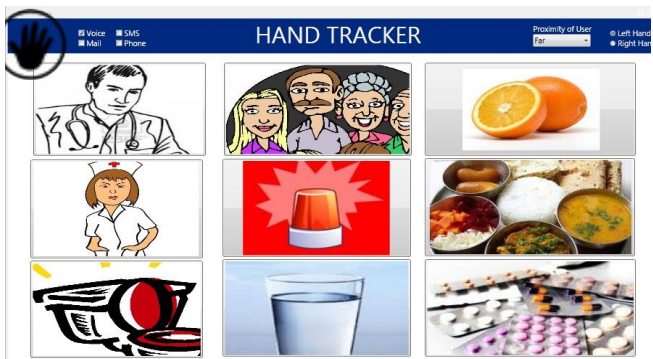

Figure. 2. Graphical user interface of the patient care system.

## IV. METHODOLOGY

*Choose primary hand*

The process starts with extracting the skeleton of the user positioned in front of the Kinect. HCI applications often differentiate between left and right hands based on several parameters and accept gestures from that particular hand. The 'primary' hand is chosen based on proximity or activity. For meaningful interpretation of the hand actions, the coordinates have to be translated to screen coordinates using appropriate scaling factors. Techniques for dynamic scaling and centering frame with respect to hands are discussed in below section. The user is required to select the radio button indicating whether to track the left hand or the right hand. By default, left hand tracking is chosen. Before tracking the hand efficiently, we first need to identify which of the two hands the user is using to interact with the system. We tried to associate natural cues that help us identify the primary hand.

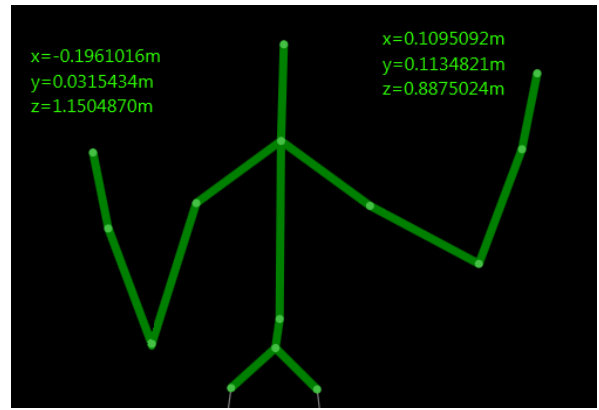

Figure 3. Skeleton extracted from Kinect depicting the nearness of the right hand when compared to the left hand. The image is flipped horizontally.

### A. Based on Nearness

While interacting with the screen it is very common for the user to face it and protrude the hand towards the screen. Thus proximity to screen is an excellent indicator to which is the primary hand to be tracked. In order to prevent erratic switching between the hands in case of hands being close together, we only switch the hand when the difference in distance exceeds a threshold. An illustrative diagram is presented in figure 3 for the same.

### B. Based on Detected Activity

Sometimes the user may not be in a position to face the screen and interact with the system. This is common when patient is in a hospital. Often when interacting using the hand, only that hand is active in motion while the other hand is mostly at rest. Thus activity of the hand is another indicator of the primary hand.

However in order to prevent erratic behavior due to involuntary hand movement, the history of activity is considered. Further in order to enable switching of hands, the history must have lower priority to the present activity. The hand with higher activity will be chosen at each frame. Thus we can define activity as-

$$A = A/2 + |d_{pos}| \qquad (1)$$

where $d_{pos}$ is the change in position of each hand and $A$ is the activity which is initialized to 0. Various techniques, which take advantage of the depth information, are used to ensure that hand tracking and mapping is effective and consistent irrespective of the proximity of user, switching of hand currently being tracked or external noise. The system does not employ special markers on the hands for tracking nor does it use any complex and computation-heavy image processing techniques in order to attain high frame rates. It involves capturing of the skeleton of the user, followed by extracting position of hands and setting cursor position accordingly in every frame. The detected hand gestures can be many depending on the patients' condition. Currently, pointing at an option and holding the pose for a couple of seconds chooses that option.

*Translating the frame of reference*

Another aspect of tracking the hand is translating the user's frame of hand movement to the screen, such that there is minimum convenience for the user and maximum usability. We define a minimum distance in x and y dimensions, that the user's hands will have to move in order to cover all the points on the screen from left to right and top to bottom. In order to facilitate this we considered following translation schemes.

### A. Dynamic Scaling of Frame by Distance

A major requirement of the system is the minimal movement of hands to control the cursor irrespective of the distance between the Kinect sensor and the user. We have to set a fixed span of length that the hand will have to move to point at any pixel on the screen. We employed a dynamic scaling technique to translate image coordinates of the hands $(x_i, y_i)$ to target position in the screen $(x_s, y_s)$. The multiplying factor will increase with the distance between the user and the sensor (depth). The depth information provided by the Kinect sensor is used in this case. The linear scaling factors in x and y dimensions ($\alpha_x$ and $\alpha_y$) at a particular depth (z), is given by eq. (2) and (3).

$$\alpha_x = \frac{z \sin(\theta_h/2)}{x_{span}/2} \frac{x_{smax}}{x_{imax}} \quad (2)$$

$$\alpha_y = \frac{z \sin(\theta_v/2)}{y_{span}/2} \frac{y_{smax}}{y_{imax}} \quad (3)$$

where $(x_{smax}, y_{smax})$ is the screen resolution in pixels, $(x_{imax}, y_{imax})$ is the dimension of the image, $(x_{span}, y_{span})$ is the maximum real-world distance that the hand will have to move in both dimensions to be able to point at all pixels on the screen, $\theta_h$ and $\theta_v$ are the horizontal and vertical field of view angles of the sensor respectively. The translated target coordinates are calculated as:

$$x_s = \alpha_x x_i \quad (4)$$
$$y_s = \alpha_y y_i \quad (5)$$

where $(x_i, y_i)$ is the pixel coordinates of the hand in the image and $(x_s, y_s)$ is the linearly translated cursor coordinates on the screen with origin positioned at the centre of the image and screen respectively.

### B. Focus the Center of Frame on the Hand Position

Due to the nature of 3D projection on a 2D image, there is a natural focus on the center of the camera frame, making out of focus/off-center positions cumbersome. Thus when the hand being tracked is present in different regions of the camera frame, the user will perceive a difference in the tracking calibration without a perceived spatial difference. In order to compensate for that, the frame of translation from 3D to projection on display uses the last tracked hand position as its reference/focus point. Thus the projection using the modified focus point is given by-

$$x_{s.new} = x_{s.old} + \alpha_x (x_{i.new} - x_{i.old}) \quad (6)$$
$$y_{s.new} = y_{s.old} + \alpha_y (y_{i.new} - y_{i.old}) \quad (7)$$

The new subscripts represent the current calculated values and the old subscript represents the values from the previous frame.

### C. Kinetic Shift of Detected Position

If the projection area is very large compared to the comfortable movement range of the user's hands, a concept similar to kinetic scrolling is used. In the hand tracking scenario, it is not feasible to use the touch system's select move release model. However inspiring from the idea of translating large velocity to intent of travelling a larger distance, we define an offset $K_x$, $K_y$ that is kinetic offset. This offset is incremented each time a large velocity is detected in a particular direction. It is decremented by motion in the opposite direction. The variation of $\Delta K_i$ vs velocity is an exponential distribution. The constants of the distribution can be set based on the user's requirement to traverse small areas quickly as compared to traversing larger areas.

*Detecting intent of action*

The next step in the process of hand tracking is detecting any intent of action by the user. In our experiment, we have considered a single possible action though it is possible to assign to different actions to different schemes of validation as they are independent [13]. Further intents are best recognized in the camera frame before project, as the projection will limit the flexibility of actions recognizable.

**Delay Based Selection:** *Intent can be detected as a persistent pause in a particular region with a certain standard deviation of error. This is the most naive method where a selection can be assumed when the user pauses for a considerable amount of time indicating a possible action.*

*Fist Detection:* A clenched fist is detected by segmentation of hand from the surrounding. This is done by obtaining the skeleton to track using the Kinect and forming an ROI around the hand. This image is segmented using one of the several methods as discussed in [18]. To detect a clenched fist, absence of fingers is detected. Fingers will make the segmented image non-convex. Thus convexity of segmented image is an indicator of a clenched fist and suitable action can be taken.

*Custom Gesture:* The possibility of using custom gestures is limitless with the constraint that it must complete processing in real time. Several alternatives have been presented in [19-21] etc. We tested simple gestures like detecting swirling finger, clasping both hands, head nod among others. In order to allow for real time processing, a state based model of updating states was used and not taking action, based on a micro-action observed. Further processing frames are differ when high velocity motion is detected.

## V. WORKING OF THE SYSTEM

Here we define screen pointer tracking as the process of tracking the position of hand in 3D space and translating the motion into corresponding actions. This enables interaction of the user with the system. An option is selected when the user points to the object for more than a predetermined period of time. This happens in following manner:

*A. Detecting a skeleton:* Microsoft Kinect API provides methods to extract the skeleton of a number of people visible in the camera view [22]. The information of the first tracked skeleton is accessed by calling the functions and data is stored in the variable Skelton.

*B. Extracting position of the hands:* The absolute position of the right and left hands has to be extracted from the skeleton information. The coordinates of the hands is then used to set the position of the hand cursor or the mouse pointer. The if-condition check which hand is chosen in the user interface and assigns the selected hand. This is shown in figure 3.

*C. Scaling and position setting of cursor*: This function maps the co-ordinates of the selected hand to the screen using a scaling factor. 1366x768 is the resolution of the canvas and suitable scaling factors are multiplied to get the right resolution of hand action to mouse movement translation. In this case, a scaling factor of 3 was applied for far tracking, while near tracking has scaling factor of around 1.8.

*D. Detecting a Click*: In the application, a user can click on his choice by pointing to the desired choice and holding it still for 2 seconds. The same function is written for every button and it is invoked every time the mouse is placed over the button. There is a global count associated with every button and this count is incremented on every invocation. Once this count crosses the threshold number of frames continuously without clicking on any other button, a message is displayed and the other actions are performed as per the user's requirements.

An alternate way to choose options was developed in an attempt to increase the success rate of click detection. A more intuitive and easily detectable gesture was thought of a gesture where both hands would be brought together in a hand clasping formation was chosen. The real-world distance between the right hand and the left hand was calculated. A 'click' would occur when this distance was below a predefined threshold value, represented by variable 'THRESHOLD'. The method resulted in a tremendous increase in accuracy of click detection because of the minimal errors in distance calculations. However, the downside of this method was that both hands had to be utilised to make a gesture of clicking. For our application of patient assistance, this idea was infeasible as we cannot expect the patients to use both hands.

*E. Channels of communication*: Different channels were used for communication.

**Email:** System send emails to the target email address using SMTP protocol. A dummy email Id was created for system.

**Phone:** To facilitate phone and SMS communication, an account at Twilio [23] was setup. It initiates a call from the Twilio number to the target number and speaks out the message stored as an XML file at a backend web location. Each option has its corresponding XML file stored in the website.

**SMS:** SMS sent using the Twilio service [24]. A new instance of the Twilio client is initiated by entering the user PIN and the password. The system stores the target phone number. Note that the users' phone number is unique and is assigned by Twilio service. In this case, it is +14692083448.

**Voice:** The voice messages are generated using the Google Text to Speech utility [25]. The nine messages are generated and stored in a folder.

## VI. EXPERIMENTAL RESULTS

This hospital patient assistance application was developed using the Kinect Development API in Microsoft Visual Studio. The application was tested on a 2.93 GHz Intel® i7 CPU system using a Microsoft Kinect sensor with color images of dimension 640x480 pixels at a frame rate of 30 fps. The application initializes by scanning for the presence of skeletons in the frame. It tracks the hand as specified and looks for hand gestures indicating selection of options, while continuously translating hand position to cursor position on the screen. When an option is selected, the application communicates with the concerned people via the desired modes by invoking a background process which sends communication request packets over the internet. The user is free to proceed making further choices right away. This ensures the smooth and uninterrupted flow of the program.

Figure 4 shows the processing time for each frame where a spike is observed at the 28th frame. This is because of overheads of invoking background processes to execute communication requests when the user selects an option on the screen. HTTP POST requests are dispatched to connect to the service providers of each mode of communication. It is observed that the average processing time (indicated by the red horizontal line) is 37.7ms. This translates to an effective frame rate of 26.52 fps. As said above, both hands can be used if configured and system will choose the correct hand form both hand.

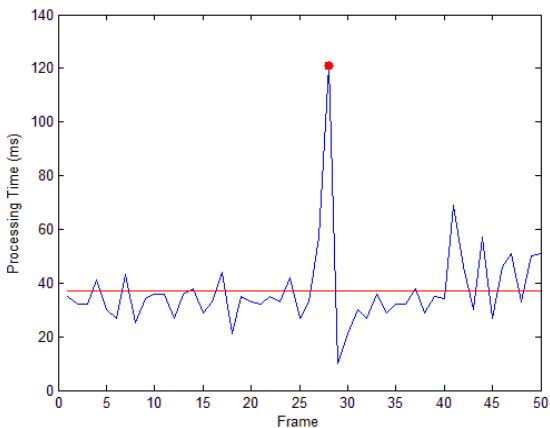

Fig. 4. Performance evaluation of the system.

The potential cursor positions of the left and right hands when they are tracked can be over nurse and fruit option. However, the fruits option would selected over the nurse option since the right hand is closer to the Kinect than the left hand. The same as been illustrated in figure 3. It can be clearly seen, based on the hand coordinates, that the right hand is closer to the sensor than the left hand. This is the reason why the option pointed to by the right hand is selected.

## VII. CONCLUSION

This paper purposes a patient-care system which is an advanced version of our last system [26]. The application consists of a GUI with a 3x3 grid of options to be selected by a patient in order to gather the attention of the concerned people using only hand gestures. The grid on the screen displays different options which can be selected by the user to summon the corresponding parties. When an option is selected the application provides options for multiple modes of communication. The salient features of the program are its customizable communication modes, proximity modes and control modes. Further, these settings can be manipulated on the fly by the simple check/uncheck the checkboxes/radio buttons provided in the user interface. The process continues till the user/caretaker closes the application window. The system has been tested thoroughly for robustness and users were happy with easy usage and response time. We would like to do further improvement based on feedback of people in future.